\documentclass[12pt,a4paper]{article}
\usepackage{jheppub}
\usepackage{subfigure}

\newcommand{\beq}{\begin{eqnarray}}
\newcommand{\eeq}{\end{eqnarray}}
\newcommand{\non}{\nonumber\\}

\newcommand{\Tr}{{\rm Tr}}
\newcommand{\diag}{{\rm diag}}

\title{On the Spectrum of Direct Gaugino Mediation}
\author{Roberto Auzzi,}
\author{Amit Giveon,}
\author{Sven Bjarke Gudnason}
\author{and Tomer Shacham}
\affiliation{Racah Institute of Physics, \\ The Hebrew University,
  \\ Jerusalem 91904, Israel}
\emailAdd{auzzi(at)phys.huji.ac.il}
\emailAdd{giveon(at)phys.huji.ac.il}
\emailAdd{gudnason(at)phys.huji.ac.il}
\emailAdd{tomer.shacham(at)phys.huji.ac.il}
\abstract{
In direct gauge mediation,
the gaugino masses are anomalously small,
giving rise to a split SUSY spectrum.
Here we investigate the superpartner spectrum
in a minimal version of ``direct {\it gaugino} mediation.''
We find that the sfermion masses are comparable to those of the
gauginos -- even in the hybrid gaugino-gauge mediation regime --
if the messenger scale is sufficiently small. }
\keywords{Supersymmetry Phenomenology}

\begin{document}
\maketitle

\section{Introduction}
If supersymmetry (SUSY) provides an explanation to the hierarchy
problem, it should be broken dynamically.
A class of such models -- ``direct gauge mediation'' -- is obtained
by embedding the Minimal Supersymmetric Standard Model (MSSM) group in the
flavor symmetry of (deformed) SQCD.
However, in such theories, the gaugino masses generically vanish to
leading order in SUSY breaking \cite{Komargodski:2009jf}. Consequently,
the sfermions are very heavy,\footnote{For a recent study of these split gauge mediated models see \cite{Shirai:2010rr}.}
and significant fine tuning of the Higgs mass is required.

The main purpose of this note is to investigate models
of SUSY breaking and its mediation to the MSSM, which have a simple,
generic dynamical origin, but nevertheless lead to
a sufficiently degenerate superpartner spectrum.
Models of ``direct gaugino mediation'' \cite{Green:2010ww} provide
such a class.
These models have a simple low-energy
effective description, which allows perturbative computations.

In this work, we compute the soft masses in the minimal version
of direct gaugino mediation. The setting of the model is presented
in section \ref{sec:setup}. It is a simple generalization of
``minimal gaugino-gauge mediation'' \cite{Cheng:2001an,Csaki:2001em},
whose sparticle spectrum was studied in detail in \cite{Auzzi:2010mb};
the messenger sector is a more general one.
For completeness, in section \ref{sec:softmasses}, we compute the soft
masses for a general messenger sector.
Then, in section \ref{sec:dynmodel}, we restrict to the particular
subclass of models
providing the effective theories of ``minimal direct gaugino
mediation''  \cite{Green:2010ww}, and present the soft masses in this class.
In section \ref{sec:rg}, we evaluate
the sparticle mass spectrum at the weak scale.

We find that for low scale mediation -- when the effective
SUSY-breaking scale is comparable to the messenger scale --
the gaugino masses can be sufficiently large
relative to the scalar masses,
even when the mass of the additional gauge particles
is also comparable to the messenger scale.
We also show that when the extra massive gauge particles
are much lighter than the messenger scale, one may ameliorate
the little hierarchy problem, allowing a relatively light stop
and a heavy Higgs field.
Finally, we discuss our results in section \ref{sec:conclusions}.

\section{Setup}\label{sec:setup}

\begin{figure}[!htp]
\centering{}
\includegraphics[width=0.5\linewidth]{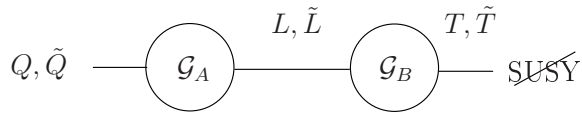}
\caption{Quiver diagram for the model.}
\label{qui}
\end{figure}

The setting \cite{Cheng:2001an,Csaki:2001em}, which is a deconstructed
version of the extra dimensional theory considered in
\cite{Kaplan:1999ac,Chacko:1999mi}, is summarized in figure
\ref{qui}.
It consists of a visible sector containing the MSSM matter fields
$Q,\tilde{Q}$ which are all charged under the gauge group
$\mathcal{G}_{A}$; a hidden sector containing messenger fields
$T_{i},\tilde{T}_{j}$ charged under a different gauge group
$\mathcal{G}_{B}$ and a pair of link fields $L,\tilde{L}$ charged
under both gauge groups.
Higgsing of the link fields breaks the symmetry
$\mathcal{G}_{A}\times\mathcal{G}_{B}$ to the diagonal
$\mathcal{G}_{\textrm{SM}}$, which is identified with the MSSM gauge
group.
The messenger fields are coupled by a superpotential to an F-term
SUSY-breaking spurion $S$, whose $\theta^{2}$ component attains a
SUSY-breaking VEV.

The group $\mathcal{G}_{A}$ consists of
  $SU(3) \times SU(2) \times U(1)$
(which we shall often think of as a subgroup of $SU(5)$), while
 we take $\mathcal{G}_{B}$ to be $SU(5)$.\footnote{This case allows perturbative unification in
a large subclass of the general models
presented in section \ref{sec:softmasses},
though it is hard to achieve unification in models
that have simple, dynamical realizations.
In section \ref{sec123} we shall also discuss the case
$\mathcal{G}_B=SU(3) \times SU(2) \times U(1)$.}
 The link fields $(L,\tilde{L})$ are chosen in the
$({\bf 5},\bar{\bf 5})$
 and $(\bar{\bf 5},{\bf 5})$ representations, respectively.
 An extra field $A$, which is an adjoint of $\mathcal{G}_B$,
 as well as an extra singlet $K$
 should be added to give masses
 to all the link field components.
 The superpotential reads:
 \begin{equation}
\mathcal {W}_{\rm link} =
\kappa_1 \,  \Tr (L A \tilde{L})
+ \kappa_2 \, K  \,  \left( \frac{\Tr (L \tilde{L})}{5} -v^2 \right) \, .
\label{superpot}
\end{equation}
The link fields obtain a VEV proportional to the $5 \times 5$
identity matrix:
 \beq
 \langle L \rangle = \langle \tilde{L} \rangle = v  \, {\bf 1}_5 \, .
 \eeq
The MSSM gauge couplings $g_k$ are related to the gauge couplings of
the unbroken theory (i.e.~before spontaneous symmetry breaking) as
follows:
\beq
\frac{1}{g_k^2}= \frac{1}{g_{A_k}^2} + \frac{1}{g_B^2} \, , \label{coupling}
\eeq
 where $k=1,2,3$ corresponds respectively to the $U(1)$, $SU(2)$ and
 $SU(3)$ gauge group,
 while the coupling $g_B$ is $SU(5)$ invariant (i.e.~the couplings of
 the $SU(3)$, $SU(2)$ and $U(1)$ subgroups are all identical).
A linear combination of the $A,B$ gauge multiplets gets a mass due to
 the super Higgs mechanism:
\beq
m_{v_k}^2 =2 v^2 (g_{A_k}^2 + g_B^2) \, . \label{massetta}
\eeq
The imaginary  part of $l_k=(L_k-\tilde{L_k})/\sqrt{2}$ is eaten via the  Higgs mechanism;
the real part of $l_k$ corresponds to the scalar in the massive vector multiplet
with mass $m_{v_k}$.

 In \cite{Auzzi:2010mb}, the sfermion soft masses have been computed at
 two loops in the case of a minimal messenger sector; in this note we
 will extend the computation to the case of a more generic weakly
 coupled messenger sector, which we will discuss in detail in the
 next section, see eq.~\eqref{genericamente}.
 A specific example motivated by the dynamical realization proposed in
 \cite{Green:2010ww} is further studied in more detail.

The two-loop calculation is not a good approximation when
the VEV $v$ is much smaller than the messenger scale $\Omega$.
In this limit, $v \ll \Omega$, the two-loop sfermion mass is
negligible; the leading contribution then comes from three loops.
An approximate computation of these contributions was performed in
\cite{DeSimone:2008gm}.
{}For the concrete example discussed in section \ref{sec:dynmodel},
we will compare the spectrum in the hybrid regime
(i.e.~$v$ of the order of $\Omega$) to the spectrum in the
limit $v \ll \Omega$.

\section{Soft masses}\label{sec:softmasses}

Let us consider the following weakly-coupled sector of $N$ messenger
pairs coupled to a SUSY-breaking F-term spurion $S$
and a D-term spurion $V$ \cite{marques,Dumitrescu:2010ha}:
\begin{align}
\mathcal{L} &= \int d^4 \theta \, \left( T_i^\dagger ( \delta_{ij} + V
\tilde{\lambda}_{ij} ) T_j
+ \tilde{T}_i^\dagger ( \delta_{ij} + V \tilde{\lambda}_{ij} ) \tilde{T}_j
\right) \non
&\phantom{=\ }
+ \int d^2 \theta\,\,  \tilde{T}_{i}\left(S \,
\lambda_{ij}+m_{ij}\right)T_{j} + {\rm c.c.}~,
\label{genericamente}
\end{align}
where $i,j=1,\ldots,N$.
We use the basis in which the matrix $m$ is diagonal with real
eigenvalues $m_i$; with this choice, by requiring messenger parity and
CP conservation, the matrix $\lambda$ is real and symmetric.
The matrix $\tilde{\lambda}$ must be Hermitian. In order to avoid a
non-zero messenger supertrace (and hence loosing calculability
\cite{Poppitz:1996xw}), we require $\Tr \, \tilde{\lambda} = 0$.
The F and D-term spurions acquire the expectation values
\beq
S |_{\theta^2} = f \, , \qquad V |_{\theta^4} = D \, ,
\eeq
which we take to be real.
The fermions in the messenger sector $(\psi_{T_i},\psi_{\tilde{T}_i})$
have Dirac masses $m_i$ while the complex scalars
$(T_i,\tilde{T}_i^*)$ have the following mass-squared matrix
\beq
\mathcal{M} = \left(\begin{array}{cc}
m^2-D \tilde{\lambda}  & -f \lambda \\
- f \lambda &  m^2 - D \tilde{\lambda} \\
\end{array}\right) \, .
\eeq
In the basis $(T_{-i},T_{+i})$, where
$T_{\pm i}\equiv (T_{i} \pm \tilde{T}_{i}^*)/\sqrt{2}$, the matrix
$\mathcal{M}$ reads:
\begin{equation}
\left(\begin{array}{cc}
m^{2} + f \lambda - D \tilde{\lambda} & 0\\
0 & m^{2} - f \lambda - D \tilde{\lambda}   \end{array}\right)
\equiv\left(\begin{array}{cc}
\mathcal{M}_{+} & 0\\
0 & \mathcal{M}_{-}\end{array}\right).
\end{equation}
Let us denote by $U_{\pm}$ the unitary matrices which diagonalize $\mathcal{M}_{\pm}$:
\beq
U_{\pm}^\dagger   \mathcal{M}_{\pm}  U_{\pm} = \diag\, (m^2_{\pm 1}, \ldots , m^2_{\pm N}) \, .
\label{uhu}
\eeq
The one-loop gaugino masses due to gauge mediation are
\cite{marques,Dumitrescu:2010ha}:
\beq
M_{\tilde{g}_k} = n_k \frac{\alpha_k}{4 \pi} \tilde{G} \, , \qquad
\tilde{G}=
 \sum_{\pm,i,j}  \mp (U^\dagger_\pm)_{i j} (U_{\pm})_{j i}
m_j  \frac{m^2_{\pm i}  \log \frac{m_{\pm i}^2}{m_j^2} }{m_{\pm i}^2-m_j^2}   \, ,
\label{gauma}
\eeq
where $\alpha_k=g_k^2/(4 \pi)$,
$k=1,2,3$ labels the gauge groups $U(1),SU(2),SU(3)$, respectively,
and $n_k$ is the Dynkin index of a \emph{single} messenger pair with
respect to the corresponding gauge group.
In the case of the gaugino-gauge mediation setting discussed in
section \ref{sec:setup}, $g_k$ is the
effective low-energy gauge coupling defined in
eq.~(\ref{coupling}), which corresponds to the unbroken combination of
the groups $\mathcal{G}_{A_k}$ and $\mathcal{G}_B$.

Let us now review the computation of the sfermion masses in
gauge mediation \cite{Dumitrescu:2010ha} in the case of an $SU(n)$
gauge group.
It is convenient to use the global $SU(n)$ current multiplet formalism
of \cite{Meade:2008wd}.
The symmetry current $j_\mu^a$ is embedded in a real superfield
$\mathcal{J}^a$, which also contains a scalar $J^a$ and a spinor
$j_\alpha^a$, where $a=1,\ldots,n^2-1$ is the adjoint index. The
functions $C_i(x)$ parametrize the correlators as follows:
\begin{eqnarray}
\langle J^a(x)J^b(0)\rangle &\equiv& C_{0}(x) \delta^{ab} \, , \non
\langle j_{\alpha}^a(x)j_{\dot{\beta}}^{*b}(0)\rangle &\equiv&
  -i\sigma_{\alpha\dot{\beta}}^{\mu}\partial_{\mu}C_{1/2}(x)
  \delta^{ab} \, ,\\
\langle j_{\mu}^a(x)j_{\nu}^b(0)\rangle &\equiv&
  \left(\eta_{\mu\nu}\partial^2 - \partial_{\mu}\partial_{\nu}\right)
  C_{1}(x) \delta^{ab} \, . \nonumber
\end{eqnarray}
In the weakly coupled setting that we consider, the components of
$\mathcal{J}^a$ can be written explicitly
\cite{Buican:2008ws,marques}:
\begin{eqnarray}
J^a &=& T_{i}^{*} t^a T_{i}
 - \tilde{ T_{i}}^{*} t^a \tilde{T}_{i} \, , \non
j_{\alpha}^a &=& -\sqrt{2}i\left(T_{i}^{*} t^a \psi_{T_{i}\alpha}
  - \tilde{T_{i}}^* t^a \psi_{\tilde{T_i}\alpha}\right)  \, , \\
j_{\mu}^a &=& i\left(T_{i} t^a \partial_{\mu} T_{i}^{*}
  -T_{i}^{*} t^a \partial_{\mu}T_{i}
  -\tilde{T}_{i} t^a \partial_{\mu}\tilde{T}_{i}^{*}
  +\tilde{T}_{i}^{*} t^a \partial_{\mu}\tilde{T}_{i} \right)
  +\psi_{T_i}\sigma_{\mu} t^a \psi_{T_i}^*
  -\psi_{\tilde{T}_i}\sigma_{\mu} t^a \psi_{\tilde{T}_i}^* \, ,  \nonumber
\end{eqnarray}
where $t^a$ are the generators of $SU(n)$.
In momentum space, the correlators are:
\begin{align}
\tilde{C}_{0}\left(p\right) &=
\sum_{\pm, i,j} \left(U_{\pm}^{\dagger}U_{\mp}\right)_{ij}    \left(U_{\mp}^{\dagger} U_{\pm}\right)_{ji}
 I(p,m_{\pm  i},m_{\mp  j})   \, ,     \nonumber \\
\tilde{C}_{1/2}\left(p\right) &=
\frac{1}{p^{2}} \sum_{\pm , i}
\left(J(m_{\pm i})-J(m_{i})\right) \nonumber \\
&+\frac{1}{p^2}\sum_{\pm,i,j}  ( U_{\pm}^{\dagger}  )_{ij}  (U_\pm)_{ji }
\left(p^{2}+m_{\pm i}^{2}-m_{j}^{2}\right)I(p,m_{\pm i},m_{j})  \, ,\nonumber \\
\tilde{C}_{1}\left(p\right) &=
 \frac{1}{3p^{2}}
   \sum_{\pm , i}  \left( \left(p^{2}+4m_{\pm  i}^{2}\right)
     I(p,m_{\pm i},m_{\pm i})+4J(m_{\pm i})\right) \nonumber \\
 &+  \frac{4}{3p^{2}} \sum_i \left(  \left(
  p^{2}-2m_{i}^{2}\right)I(p,m_{i},m_{i}) -2 J(m_{i}) \right)  \, ,
\end{align}
where
\begin{align}
I(p,m_{1},m_{2}) \equiv \int  \frac{d^4 q}{(2  \pi)^4}
\frac{1}{\left(\left(p+q\right)^{2}+m_{1}^{2}\right)\left(q^{2}+m_{2}^{2}\right)}
\, , \quad
J(m) \equiv \int  \frac{d^4 q}{(2  \pi)^4}
\frac{1}{\left(q^{2}+m^{2}\right)}  \, . \nonumber
\end{align}
The sfermion soft masses are then given by:
\begin{equation}
m_{\tilde{f}}^{2} = - 8\pi^2 \sum_{k=1}^3 \alpha_k^2 C_{\tilde{f},k} n_k
  \int \frac{d^4 p}{(2\pi)^4}\frac{f_k(p^2)}{p^{2}}
  \left(\tilde{C}_{0}(p)-4\tilde{C}_{1/2}(p)
  +3\tilde{C}_{1}(p)\right) \, ,
\label{eq:Sfermion Mass}
\end{equation}
where the index $k$ sums over the gauge groups $U(1)$, $SU(2)$ and
$SU(3)$; $C_{\tilde{f},k}$ is the quadratic Casimir of the sfermion
$\tilde{f}$ and $n_k$ is the Dynkin index of a \emph{single} pair of
messengers. Finally, the function $f_k(p^2)$ 
depends on the gauge group as well as the specific model under
consideration. In the case of gauge mediation it is simply
$f_k(p^2) = 1$ for all $k$.
A direct calculation \cite{Dumitrescu:2010ha} using the two-loop
integrals in \cite{Martin:1996zb} and setting $f_k(p^2)=1$ gives rise
to:
\begin{align}
m^2_{\tilde{f}} =
2 \sum_{k=1}^3 \left( \frac{\alpha_k}{4 \pi} \right)^2 \,
C_{\tilde{f}, k}   n_k
\Bigg\{ &\sum_{\pm , i} \left( m_{\pm  i}^2 \log  m_{\pm i}^2
  -  m_{i}^2 \log  m_{i}^2 \right)
\label{fermama1} \\
&+  \frac{1}{2} \sum_{\pm,i,j} (U_{\pm}^\dagger U_{\mp})_{ij}
(U^\dagger_{\mp}  U_{\pm})_{ji}  m_{\pm  i}^2 {\rm Li}_2 \left(
1-\frac{m_{\mp  j}^2}{m_{\pm i}^2} \right) \non
&- 2 \sum_{\pm,i,j}(U_{\pm}^\dagger)_{ij} (U_\pm)_{ji}  m_{\pm i}^2  {\rm Li}_2
\left( 1-\frac{m_{j}^2}{m_{\pm  i}^2} \right)
 \Bigg\} \ , \nonumber
\end{align}
where the dilogarithm is defined by
${\rm Li}_2(x)=-\int_0^1\frac{dt}{t}\log(1-xt)$.

In order to compute the two-loop sfermion masses in the gaugino-gauge
mediation setting discussed in section \ref{sec:setup},
we should use the momentum dependent function found in
\cite{McGarrie,Auzzi:2010mb,Sudano:2010vt}:
\beq
f_k(p^2) = \left( \frac{m_{v_k}^2}{p^2+m_{v_k}^2} \right)^2 \, .
\eeq
Using the two-loop integrals in \cite{vanderBij:1983bw,Martin:1996zb},
an explicit calculation gives:
\beq
m_{\tilde{f}}^2 &=& 2\sum_{k=1,2,3}
\left( \frac{\alpha_k}{4 \pi} \right)^2  C_{\tilde{f}, k} n_k
\tilde{S}_k \, ,  \label{fermama2} \\
\tilde{S}_k &=&
 \sum_{\pm,i,j} (U_{\pm}^\dagger U_{\mp})_{ij} (U_{\mp}^\dagger U_{\pm})_{ji}  a_{\pm ij}^k
+   \sum_{\pm,i,j} (U_{\pm}^\dagger)_{ij}  (U_{\pm})_{ji} b_{\pm ij}^k
 +  \sum_{\pm,i} c_{\pm i}^k  \, ,
\eeq
where we have defined
\beq
a_{\pm i j}^k &=&\frac{1}{2} m_{\pm i  }^2 \left(
{\rm Li}_2 \left( 1- \frac{m_{\mp j}^2}{m_{\pm i}^2}  \right)
- h \left( \frac{m_{v_k}^2}{m_{\pm i}^2}, \frac{m_{\mp j}^2}{m_{\pm i}^2} \right)
\right) \, , \non
b_{\pm i j}^k &=& \left( \frac{2(m_j^2-m_{\pm i}^2) }{m_{v_k}^2} +1  \right) \left(m_j^2
h \left( \frac{m_{v_k}^2}{m_j^2}, \frac{m_{\pm i}^2}{m_j^2} \right)
+ m_{\pm i}^2
h \left( \frac{m_{v_k}^2}{m_{\pm i}^2}, \frac{m_{j}^2}{m_{\pm i}^2} \right)
 \right) \non
&&+ (m_j^2-m_{\pm i}^2)
 \left( \frac{ m_j^2 }{m_{v_k}^2} \log^2 \left( \frac{m_j^2}{m_{\pm i}^2} \right)
 +   h \left( \frac{m_{\pm i}^2}{m_{v_k}^2} , \frac{m_j^2}{m_{v_k}^2} \right)
 \right) \non
&&+2 \left( \frac{(m_j^2-m_{\pm i}^2)^2 }{m_{v_k}^2} -m_{\pm i}^2 \right)
 {\rm Li}_2 \left( 1- \frac{m_j^2}{m_{\pm i}^2}\right) \, , \\
c_{\pm i}^k &=& m_{\pm i}^2 \left(
-\frac{2 m_{\pm i}^2}{m_{v_k}^2} +  \log m_{\pm i}^2 +
  h \left( \frac{m_{\pm i}^2}{m_{v_k}^2}, \frac{m_{\pm i}^2}{m_{v_k}^2} \right)
 + \left( \frac{4 m_{\pm i}^2 }{  m_{v_k}^2} -\frac{1}{2} \right)
 h \left(\frac{m_{v_k}^2}{m_{\pm i}^2},1 \right)
\right) \non
&&- m_{ i}^2 \left(
-\frac{2 m_{ i}^2}{m_{v_k}^2} +  \log m_{ i}^2 +
  h \left( \frac{m_{ i}^2}{m_{v_k}^2}, \frac{m_{ i}^2}{m_{v_k}^2} \right)
 + \left( \frac{4 m_{ i}^2 }{ m_{v_k}^2}  +1 \right)
  h \left(\frac{m_{v_k}^2}{m_{ i}^2},1 \right)
\right) \, . \nonumber
\eeq
This expression generalizes the one found in \cite{Auzzi:2010mb} for a
minimal messenger sector.
The function $h$ is defined by the integral:
\begin{equation}
h(a,b)= \int_0^1 dx  \left( 1+ {\rm Li}_2 (1-\mu^2)
-\frac{\mu^2}{1-\mu^2}  \log \mu^2 \right) \, ,
\qquad \mu^2=\frac{a x + b(1-x)}{x(1-x)} \, ; \nonumber
\end{equation}
an analytical expression for $h$ can be found in
\cite{vanderBij:1983bw}.
Note that eq.~\eqref{fermama1} is recovered from
eq.~\eqref{fermama2} by taking the limit $m_{v_k}\to\infty$.

\section{A dynamical realization} \label{sec:dynmodel}

\subsection{Description of the model} \label{sec:dynmodelA}

A realization of the gaugino mediation setup described
in section \ref{sec:setup} in massive SQCD with singlets was
studied in \cite{Green:2010ww}.
Consider $SU(N_{c})$ SQCD with $N_{f}$ quark multiplets
$Q$ in the fundamental representation
as well as $N_f$ anti-fundamental multiplets $\tilde{Q}$,
which are labeled by the indices $i,j=1,\ldots,n $ and
$a,b=n+1,\ldots,N_f$, where $n=N_f-N_c$. The singlets $H^a_i$ and
$\tilde{H}^i_a$ are introduced as well.
The superpotential reads
\begin{equation}
\mathcal{W}_e=\left(\begin{array}{cc}
Q^{i} & Q^{a}\end{array}\right)
\left(\begin{array}{cc}
m_{(1)}\delta_{i}^{j} & H_{i}^{b}\\
\tilde{H}_{a}^{j} & m_{(2)}\delta_{a}^{b}\end{array}\right)
\left(\begin{array}{c}
\tilde{Q}_{j}\\
\tilde{Q}_{b}\end{array}\right) \, .
\end{equation}
The masses of the two subsets of quarks, $m_{(1)}$ and
$m_{(2)}$, are both taken to be much smaller than the confinement
scale $\Lambda$. For $N_{f}<3N_{c}$ this
theory is asymptotically free; the Seiberg dual  \cite{Seiberg:1994pq},
for $N_f>N_c+1$, is given in terms of an $SU(n)$ gauge theory.
We focus on the regime $N_f < 3 N_c /2$, where the dual theory is IR-free.
The superpotential of the dual theory is then:
\begin{equation}
\mathcal{W}_m = q M \tilde{q}
+ \Lambda H_{i}^{a} M_{a}^{i} + \Lambda\tilde{H}_{a}^{i} M_{i}^{a}
+ \Lambda \, \Tr\left(\boldsymbol{m} \, M\right) \, ,
\end{equation}
where $q,\tilde{q}$ are the $N_f$ dual quarks, $M$ is the meson field and
$\boldsymbol{m}=m_{(1)}\delta_{i}^{j}+m_{(2)}\delta_{a}^{b}$.

After integrating out the heavy states, the meson and
the dual quarks can be decomposed into the following blocks
(in flavor space, while the color indices for the dual quarks are
implicit):
\beq
M = \left(
\begin{array}{c|c}
N_{n \times n} &  \\
\hline
 & \begin{array}{c|c}
X_{n \times n} & Y_{n \times p} \\
\hline
\raisebox{-2pt}{$\tilde{Y}_{p \times n}$} &
\raisebox{-2pt}{$Z_{p \times p}$}
\end{array}
\end{array}
\right) \, , \qquad
q= \left(\begin{array}{c|c|c}
\chi_n &
\eta_n &
\rho_p
\end{array}\right)  \, ,
\qquad
\tilde{q}  =
\left(\begin{array}{c}
\tilde{\chi}_n \\ \hline
\tilde{\eta}_n \\ \hline
\tilde{\rho}_p
\end{array}\right)  \, ,
\eeq
where the subscripts denote the dimension of each block
and $p=2N_c-N_f$.
In order to cancel as many F-terms as possible, the VEVs are chosen as follows:
\beq
\chi=\tilde{\chi} = \sqrt{\Lambda m_{(1)}} {\bf 1}_n  \, , \qquad
\eta \, \tilde{\eta} = \Lambda m_{(2)} {\bf 1}_n \, ,
\eeq
while the $Z$ components are pseudo-moduli
and the other VEVs are equal to zero.
The one-loop Coleman-Weinberg potential sets $Z=0$
and $\eta=\tilde{\eta}$; the SUSY-breaking
sector is similar to the ISS model \cite{Intriligator:2006dd}.

The fields $\chi, \tilde{\chi}$ are identified
with the link fields $L,\tilde{L}$
of the quiver diagram in figure \ref{qui} and
the $SU(n)_\chi$ flavor group is identified
with the gauge group $\mathcal{G}_A$.
The VEV of $\eta$ breaks the $SU(n)_\eta$ global flavor group
and the dual $SU(n)$ gauge group to a diagonal combination,
 which is identified with the gauge group $\mathcal{G}_B$.
Finally, the field $Z$ is the SUSY-breaking spurion $S$ while
 \beq T=(T_1,T_2)=(\rho,Y) \, , \qquad \tilde{T}=(\tilde{T}_1,\tilde{T}_2)
 =(\tilde{\rho},\tilde{Y}) \, ,
 \label{messi} \eeq
are the messengers.

The model has an accidental R-symmetry which  is not broken by the metastable
vacuum; for this reason the gaugino masses are zero.
This accidental symmetry should thus be broken,
e.g.~by adding a quartic deformation in the UV,
as in \cite{Giveon:2007ef}.
This deformation turns in the IR into a superpotential term
$\delta\mathcal{W} \propto \Tr \, Z^2$, giving $Z$ a non-zero VEV which
we choose to be $\langle Z \rangle = \omega \, {\bf 1}_p$.

\subsection{Soft masses} \label{sec:dynmodelB}

In this section we apply the generic results of section
\ref{sec:softmasses}
to the model briefly described in section \ref{sec:dynmodelA}.
There is a total of $2 p$ messengers which can be organized as $p$ copies of
 the two messenger pairs $(T_1,\tilde{T}_1)$ and $(T_2,\tilde{T}_2)$.
The coupling between the messengers and the spurion $S$
is given by eq.~(\ref{genericamente}) with
\begin{equation}
m = \left(\begin{array}{cc}
\omega  & \Omega   \\
\Omega & 0
\end{array}\right)  \otimes {\bf 1}_p \, , \qquad
\lambda =  \left(\begin{array}{cc}
1  & 0   \\
0 & 0
\end{array}\right) \, \otimes {\bf 1}_p \, ,
\end{equation}
where $\Omega=\sqrt{\Lambda m_{(2)}}$ and $\omega$ is the VEV of each
diagonal element of $Z$.
The VEV of the link field $L$ is $v = \sqrt{\Lambda m_{(1)}}$.
In order to satisfy the relation $N_f < 3 N_c/2$
for $n=5$ (i.e.~corresponding to an $SU(5)$ GUT),
we have to require $p \geq 6$; the values of $(n,p)=(5,6)$
correspond to $N_c=11$, $N_f=16$. A more generic expression is:
\beq
N_c=n+p \, , \qquad N_f= 2 n + p \, .
\eeq
In the following we will first consider the formal case $p=1$ which
corresponds to two messengers.
The only effect of general $p$ is an overall $p$-factor in
the soft masses, which we shall reintroduce in the end.

It is useful to pass to the basis where the masses of the fermionic
messengers are diagonal:
\beq
m = \left(\begin{array}{cc}
m_{1}  &  0 \\
0  & m_{2} \\
\end{array}\right)  \, , \qquad
\lambda =
\frac{1}{2}\mathbf{1}_2
+\frac{1}{2\sqrt{4\Omega^2 + \omega^2}}
\left(\begin{array}{cc}
-\omega & 2\Omega \\
2\Omega & \omega
\end{array}\right) \, ,
\eeq
where
\beq
m_{1,2}=\frac{1}{2} (\omega \mp \sqrt{4 \Omega^2 + \omega^2}) \, .
\eeq
The squared masses of the bosonic messengers are:
\beq
m^{2}_{\pm s}=\frac{1}{2} \left( \pm f + 2\Omega^2 + \omega^2 + (-1)^s
\sqrt{f^2+2(2\Omega^2 \pm f) \omega^2+\omega^4}
\right) \, ,
\eeq
where $s=1,2$.
The diagonalization matrices for the bosonic messenger masses in
eq.~(\ref{uhu}) are given by:
\beq
U_\pm =\left(
\begin{array}{cc}
\cos \theta_\pm & \sin \theta_\pm  \\
-\sin \theta_\pm &  \cos \theta_\pm
\end{array}
\right) \, ,
\eeq
where the rotation angles are:
\beq
\tan \theta_{\pm} =
\frac{\omega(\mp f - 4\Omega^2 - \omega^2)
+\sqrt{(4\Omega^2 + \omega^2)(f^2\pm 2f + 4\Omega^2\omega^2 + \omega^4)}}
{\pm 2f\Omega} \, .
\eeq

\begin{figure}[!htp]
\centering{}
\includegraphics[width=0.4\linewidth]{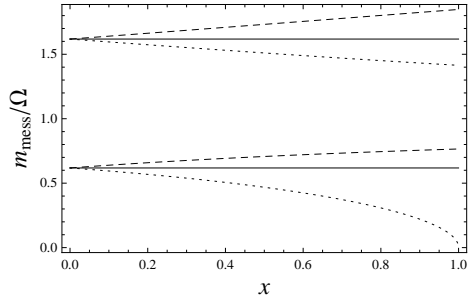}
\caption{Messenger masses as functions of $x$ in units of $\Omega$ for
  $z=1$.}
\label{messaggerifig}
\end{figure}

The following variables are introduced for convenience:
\beq
x=\frac{f}{\Omega^2} \, , \qquad
y_k=\frac{m_{v_k}}{\Omega} \, , \qquad z=\frac{\omega}{\Omega} \, .
\eeq
In the numerical examples that we will consider,
the relative difference between the parameters $y_k$, $k=1,2,3$, is
very small; hence, we will often simply denote by $y$ their average
value. Analogously, we will denote by $m_v$ the average of $m_{v_k}$.
A plot of the messenger masses as a function of $x$ for $z=1$ is shown
in figure \ref{messaggerifig}.

\begin{figure}[ht]
\begin{center}
$\begin{array}{c@{\hspace{.2in}}c@{\hspace{.2in}}c} \epsfxsize=2.5in
\epsffile{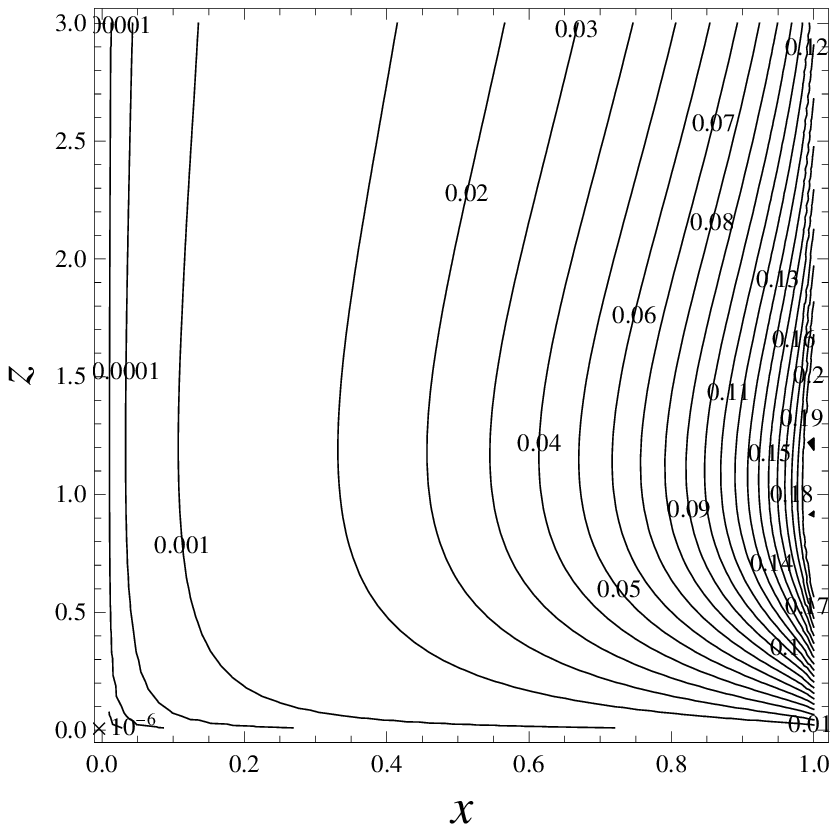}  &
     \epsfxsize=2.7in
    \epsffile{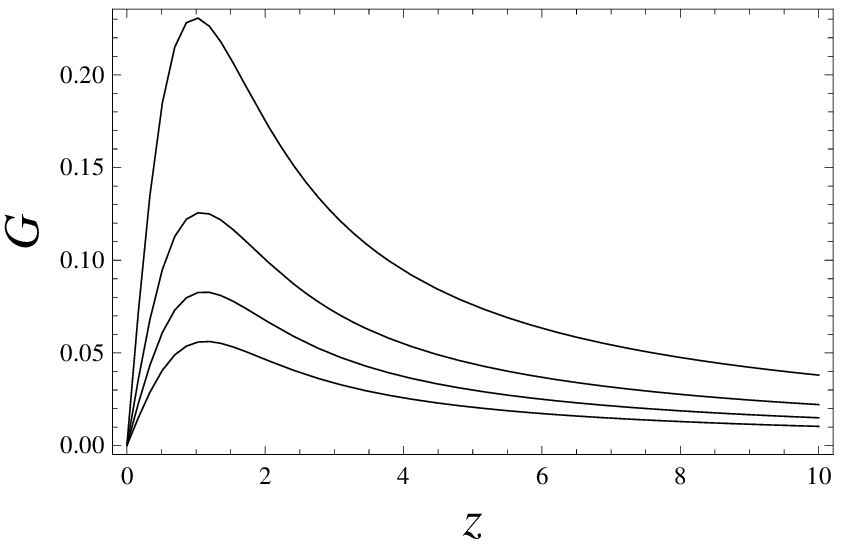}
\end{array}$
\end{center}
\caption{\footnotesize
Left panel:  The function $G(x,z)$ for the gaugino masses, as defined
in eq.~\eqref{maga}.
Right panel: The function $G(x,z)$ as a function
of $z$ with $x=1,0.9,0.8,0.7$  (from top to bottom). }
\label{gaugino}
\end{figure}

Using eq.~(\ref{gauma}), we can compute the gaugino soft masses:
\beq
M_{\tilde{g}_k}= p  \frac{\alpha_k}{4\pi}
\frac{f}{\Omega}  \, G(x,z) \, ,
\qquad G(x,z) = \frac{\Omega}{f} \tilde{G} \, , \label{maga}
\eeq
where $G$ is the function calculated for $p=1$ (i.e.~corresponding to
2 messengers)  and $\alpha_k=g_k^2/(4\pi)$,
where $g_k$ is defined in eq.~(\ref{coupling}).
The messengers are  fixed here in  the ${\bf 5}+\bar{\bf 5}$ representation  of $SU(5)$.
Notice the reinstated $p$, which renders the result valid for any $p\geq 1$.
A plot of the function $G$ is shown in figure \ref{gaugino};
for fixed $x$, we obtain the highest gaugino masses for $z \approx 1$.

\begin{figure}[ht]
\begin{center}
$\begin{array}{c@{\hspace{.2in}}c@{\hspace{.2in}}c} \epsfxsize=2.5in
\epsffile{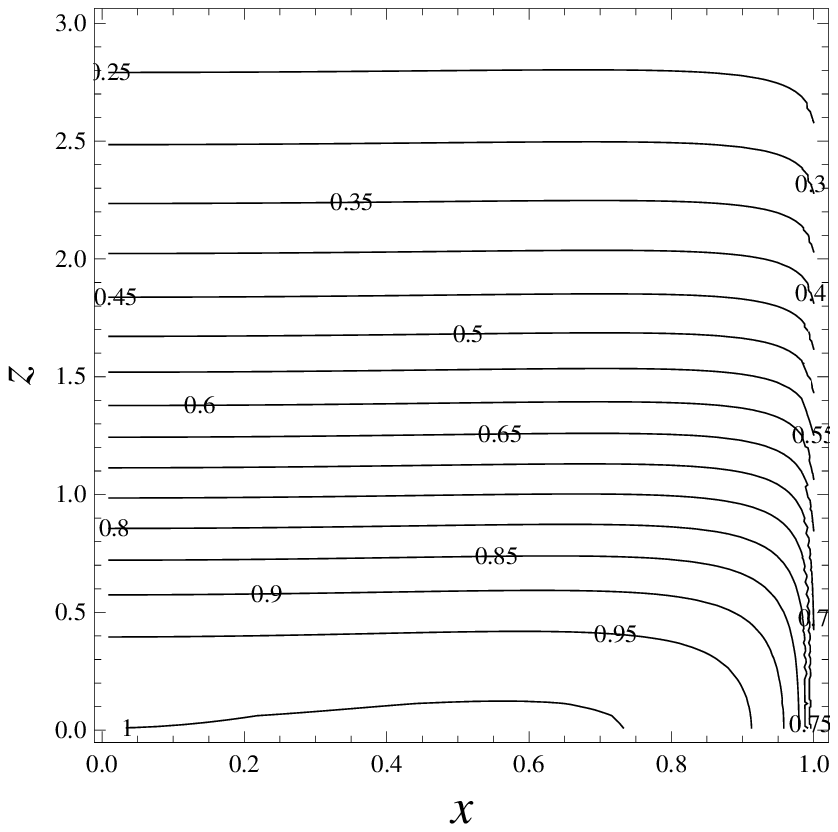}  &
     \epsfxsize=2.5in
    \epsffile{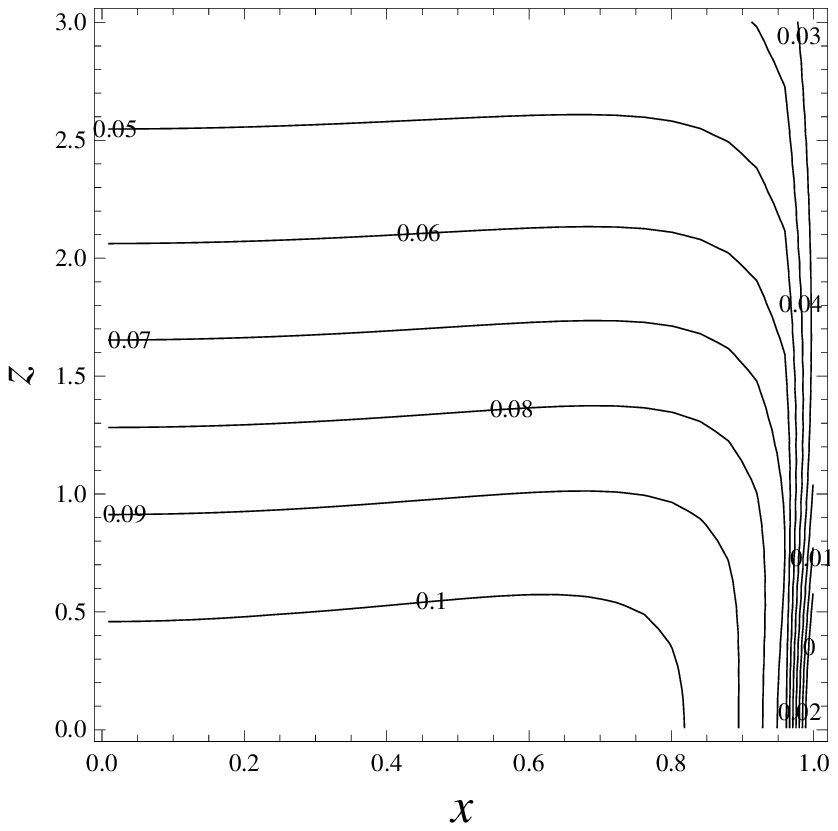}
\end{array}$
\end{center}
\caption{\footnotesize
Left panel: The function $S(x,y,z)$ for the sfermion squared masses
(as defined in eq.~\eqref{masisi}), in the limit $y\to\infty$.
Right panel: The function $S(x,1,z)$, viz.~with $y=1$.  }
\label{sferm1}
\end{figure}

In the limit $x\to 1$, the function $G(x,z)$ has the following
behavior:
\begin{equation}
\lim_{x\to 1} G(x,\omega/\Omega) \sim  \left\{ \begin{array}{cl}
0.39\, \Omega/\omega & ,\quad \omega\gg\Omega \, ,\\
0.23 & ,\quad \omega\sim \Omega  \, ,\\
0.45\, \omega/\Omega & ,\quad \omega\ll\Omega  \, .\end{array}  \right.
\end{equation}
Using eq.~\eqref{fermama2}, we obtain for the sfermion mass-squared:
\begin{align}
m^2_{\tilde{f}}=2p
\sum_k \left( \frac{\alpha_k}{4 \pi} \right)^2
 C_{\tilde{f}, k}
 \left( \frac{f}{\Omega} \right)^2 S  (x,y_k,z)  \, , \qquad
S(x,y_k,z) = \left( \frac{\Omega}{f} \right)^2 \tilde{S}_k \, .
\label{masisi}
\end{align}
The function $S$ is again calculated for the case $p=1$, while the
explicit factor of $p$ appears as the promised multiplicative
factor and hence renders the result valid for any $p\geq 1$.
The function $S$ is shown in figure \ref{sferm1} in the large $y$
limit as well as for $y=1$.
As already mentioned, in the large $y$ limit, eq.~(\ref{masisi})
reduces to the gauge mediation expression given by
eq.~(\ref{fermama1}).

\section{Weak scale spectrum}\label{sec:rg}

\subsection{Hybrid regime}

{}From figure \ref{gaugino}, one can see that the one-loop gaugino
masses are indeed highly suppressed when $x$ is small.
The regime that is more interesting phenomenologically
(i.e.~in order to avoid large sfermion masses) thus
corresponds to relatively large $x$, i.e.~roughly $0.8 < x < 1$, and
$z \approx 1$.
We are then forced to consider only low messenger scales, such as
$10^5 - 10^6 \, {\rm GeV}$.
In this regime of parameter space, the various masses of the
messengers are rather split (see figure
\ref{messaggerifig}), however, the average messenger scale is roughly
$\Omega$ in any case.

The two-loop sfermion masses computed in the previous
section can be trusted only in the regime where
the function $S(x,y,z)$ in eq.~(\ref{masisi}) is sufficiently larger than a loop
factor $\alpha/(4 \pi)\approx 0.01$.
In the model under consideration, this is usually the case for
$y \gtrsim 1$.
Examples of spectra in this regime are shown in table \ref{tab1}  (for
$p=1$) and in table \ref{tab2} (for $p=6$).
The case of $p=6$ corresponds to the minimal $N_c=11$ and $N_f=16$
embedding in the dynamical model discussed in section
\ref{sec:dynmodel}, while $p=1$ does not correspond to any known
dynamical embedding, but formally it is the minimal case
with vanishing gaugino mass at the leading order in
SUSY breaking -- the case with two messengers.
The spectra were obtained using the program SOFTSUSY
\cite{Allanach:2001kg} to solve the renormalization group (RG)
equations from the messenger scale $\Omega$ to the weak scale.
The mass splitting among the messengers is not negligible
(e.g. there is a ratio of $6$ between the mass of the heaviest and
the lightest messengers for $x=0.8$ and a ratio of $20$ for $x=0.98$),
and thus a priori there is no clear-cut scale from which it is most
appropriate to start the RG evolution.
The imprecision arising from this fact is however of the order of
higher order corrections in $\alpha_k$.
We did check though that the results are not that sensitive to this
choice in the specific examples that we have considered.
The trilinear scalar soft terms were set to zero at the messenger
scale due to an extra loop suppression.
We assume that only the gauge mediated contribution is present for
the soft masses of $H_u,H_d$ at the messenger scale while
$\mu, B \mu$ are computed by imposing electroweak symmetry breaking
as well as the value of $\tan\beta$.

\begin{table}[!htp]
\centering{}\begin{tabular}{|c|c|c|}
\hline
$\Omega$ & $ 1.7\times 10^6 $ & $5.15\times 10^5$  \tabularnewline
$x,y,z$ & 0.8,1,1 & 0.98,1.8,1  \tabularnewline
\hline
\hline
$S_{1},S_{2},S_{3}$ & 0.08,0.09,0.10 & 0.13, 0.13, 0.15 \tabularnewline
\hline
$M_{\tilde{g}_1},M_{\tilde{g}_2},M_{\tilde{g}_3}$ & 179,303,601 & 153,266,555\tabularnewline
$m_{\tilde{Q}},m_{\tilde{u}},m_{\tilde{d}}$ & 3974,3769,3748 & 1911,1821,1812 \tabularnewline
$m_{\tilde{L}},m_{\tilde{e}}$ & 1381,692 & 631,309 \tabularnewline
$\mu,B\mu$ & 1565,2011$^{2}$ & 784,991$^{2}$ \tabularnewline
\hline
\hline
$m_{\tilde{g}}$ & 941 &  818 \tabularnewline
$m{}_{\tilde{\chi}_{0}}$ & 154,308,1545,1546 & 133,263,774,778\tabularnewline
$m{}_{\tilde{\chi}_{\pm}}$ & 308,1547 & 263,779 \tabularnewline
\hline
$m_{\tilde{u}_{L}},m_{\tilde{d}_{L}}$ & 4013,4014 &  1961,1963 \tabularnewline
$m_{\tilde{u}_{R}},m_{\tilde{d}_{R}}$ & 3802,3781 & 1870,1862 \tabularnewline
$m_{\tilde{t}_{1}},m_{\tilde{t}_{2}}$ & 3408,3815 & 1688,1872 \tabularnewline
$m_{\tilde{b}_{1}},m_{\tilde{b}_{2}}$ & 3742,3814 & 1841,1870 \tabularnewline
\hline
$m_{\tilde{e}_{R}},m_{\tilde{e}_{L}},m_{\tilde{\nu}_{e}}$ & 700,1378,1376 & 317,635,630 \tabularnewline
$m_{\tilde{\tau}_{1}},m_{\tilde{\tau}_{2}},m_{\tilde{\nu}_{\tau}}$ & 679,1374,1370 & 304,635,628 \tabularnewline
\hline
$m_{h_{0}}$ & 121 & 117 \tabularnewline
$m_{H_{0}},m_{A_{0}},m_{H_{\pm}}$ & 1965,1965,1967 & 947,947,950 \tabularnewline
\hline
\end{tabular}\caption{Examples of weak scale spectra in the case of
  $p=1$ (two messengers),
with $\mu>0$, $\tan \beta = 20$ and $\alpha_B^{-1}=4$ at the scale $\Omega$.
All the masses are in GeV. $y$ is the average of $y_1$, $y_2$, $y_3$
while $S_k$ is an abbreviation for $S(x, y_k, z)$.
The input masses $(M_{\tilde{g}_k},m_{\tilde{Q},\tilde{u},\tilde{d},\tilde{L},\tilde{e}})$
are given at the messenger scale $\Omega$; $(\mu, B \mu)$ are
also evaluated at $\Omega$. The other masses in the table are the MSSM
pole masses. Here we have included only the
 two-loop contribution to the sfermion soft masses. }
\label{tab1}
\end{table}

\begin{table}[!htp]
\centering{}\begin{tabular}{|c|c|c|}
\hline
$\Omega$ & $2.6\times 10^5 $ & $ 1.65\times 10^5 $  \tabularnewline
$x,y,z$ & 0.8,1,1 & 0.98,1.8,1  \tabularnewline
\hline
\hline
$S_{1},S_{2},S_{3}$ & 0.08,0.09,0.10 & 0.13, 0.13, 0.16 \tabularnewline
\hline
$M_{\tilde{g}_1},M_{\tilde{g}_2},M_{\tilde{g}_3}$ & 159,278,598 & 288,505,1094\tabularnewline
$m_{\tilde{Q}},m_{\tilde{u}},m_{\tilde{d}}$ & 1616,1545,1538 & 1537,1469,1463 \tabularnewline
$m_{\tilde{L}},m_{\tilde{e}}$ & 517,251 & 490,237\tabularnewline
$\mu,B\mu$ & 648,833$^{2}$ & 611,858$^{2}$ \tabularnewline
\hline
\hline
$m_{\tilde{g}}$ & 849 & 1419 \tabularnewline
$m{}_{\tilde{\chi}_{0}}$ & 139,273,640,648 & 254,478,605,631\tabularnewline
$m{}_{\tilde{\chi}_{\pm}}$ & 274,648 & 478,630\tabularnewline
\hline
$m_{\tilde{u}_{L}},m_{\tilde{d}_{L}}$ & 1678,1680 &  1716,1718 \tabularnewline
$m_{\tilde{u}_{R}},m_{\tilde{d}_{R}}$ & 1606,1600 & 1651,1646 \tabularnewline
$m_{\tilde{t}_{1}},m_{\tilde{t}_{2}}$ & 1464,1610 & 1525,1660 \tabularnewline
$m_{\tilde{b}_{1}},m_{\tilde{b}_{2}}$ & 1582,1607 & 1630,1654 \tabularnewline
\hline
$m_{\tilde{e}_{R}},m_{\tilde{e}_{L}},m_{\tilde{\nu}_{e}}$ & 259,524,518 & 252,514,508 \tabularnewline
$m_{\tilde{\tau}_{1}},m_{\tilde{\tau}_{2}},m_{\tilde{\nu}_{\tau}}$ & 248,524,516 & 241,515,506 \tabularnewline
\hline
$m_{h_{0}}$ & 116 & 116 \tabularnewline
$m_{H_{0}},m_{A_{0}},m_{H_{\pm}}$ & 780,781,785 &   748,748,753 \tabularnewline
\hline
\end{tabular}\caption{Examples of weak scale spectra in the case of
  $p=6$ (six identical copies of two messengers), with $\mu>0$,  $\tan \beta = 20$ and
  $\alpha_B^{-1}=4$ at the scale $\Omega$. }
\label{tab2}
\end{table}

{}For $p=1$ we get a (mostly bino) neutralino NLSP in all of the
parameter space, while for $p=6$ both neutralino and stau NLSPs are
possible.
The neutralino NLSP is promptly decaying, because $\sqrt{f}  \lesssim
10^6 \, {\rm GeV}$.
The experimental constraints from $36 \, {\rm pb}^{-1}$ of  LHC data
discussed in \cite{CMS} are relevant in the case of neutralino NLSP,
implying that the gluino mass has to be greater than
$600 \, {\rm GeV}$. For most spectra in the tables we have been able
to put the Higgs mass near $115 \, {\rm GeV}$, whereas in a few examples it
was necessary to push it a bit in order to obtain a gluino mass above $800 \, {\rm GeV}$.

\subsection{Gaugino mediation regime}

\subsubsection{$\mathcal{G}_B=SU(5)$} \label{sususec1}

This regime is defined by $m_{v_k}\ll\Omega$ and in this limit the link
field potential due to SUSY breaking is no longer negligible
compared to the tree-level potential arising from the superpotential in
eq.~\eqref{superpot}.
{}For this reason we redefine the scale $v$ in eq.~(\ref{superpot})
to be $\Lambda_v$:
\begin{equation}
\mathcal {W}_{\rm link} =
\kappa_1 \,  \Tr (L A \tilde{L})
+  \kappa_2 \, K  \,
\left( \frac{\Tr (L \tilde{L})}{5} -\Lambda_v^2 \right) \, ,
\end{equation}
while we continue to denote by $v$ the VEV of the link field, which
is now obtained by minimizing the link field potential:
\beq
V_{\rm link}= V_F + V_D+
 m_L^2 \Tr \, (L L^\dagger+\tilde{L}^\dagger \tilde{L}) \, , \label{poto}
\eeq
where $V_F$ and $V_D$ are the supersymmetric F and D-term potentials
and $m_L$ is the gauge mediated soft mass for the link field, which in
the limit $m_{v_k} \ll \Omega$ can be computed from the gauge mediation
expression (\ref{fermama1}),(\ref{fermama2}), with $\alpha_k$ replaced by
$\alpha_B$,
\beq
m_L^2 = 2\left(\frac{\alpha_B}{4\pi}\right)^2 C_L \, p
\left(\frac{f}{\Omega}\right)^2 S(x,\infty,z) \ ,
\eeq
where $C_L=12/5$ is the Casimir of the link field.
The heavy vector still has the mass $m_{v_k}$ given by
eq.~(\ref{massetta}) while the link field scalar instead gets a mass
$m_{s_k}^2=m_{v_k}^2+2 m_L^2$.

For $m_{v_k} \ll \Omega$, the two-loop sfermion masses
computed in sections \ref{sec:softmasses} and \ref{sec:dynmodel} are negligible
and hence
the leading order sfermion masses are generated at three loops. Thus
we can use the method described in \cite{DeSimone:2008gm} to compute the
spectrum.

Let us denote by $m_v$ the average of the masses $m_{v_k}$.
It is useful to split the renormalization of the soft parameters
into two parts, $m_{v}<\mu<\Omega$ and $\mu < m_{v}$,
according to the RG scale $\mu$.
At the scale $\Omega$ the sfermion masses are taken to be zero,
while  the gaugino of the gauge group $\mathcal{G}_B$
gets a soft mass $M_{\tilde{g}, B}$, which can be computed from the
expression (\ref{gauma}, \ref{maga}), with $\alpha_k$ replaced
 by $\alpha_B$. The link field scalars $L,\tilde{L}$
also get a gauge mediated soft mass $m_L$.
In the present example,
$M_{\tilde{g}, B}$ vanishes to leading order in SUSY breaking and
hence is suppressed compared to $m_L$.

{}From the scale $\Omega$ to $m_{v}$, the renormalization group
equations for $(m^2_L,M_{\tilde{g}, B})$ are given by
\cite{martin-review}:
\beq
\frac{d \, m^2_L }{d (\log \mu)} =
-\frac{C_L}{2\pi^2} g_B^2 M_{\tilde{g},B}^2 \, , \qquad
\frac{d \, M_{\tilde{g}, B} }{d (\log \mu)}   =
\frac{b_B}{8 \pi^2} g^2_{B}  M_{\tilde{g},B } \, ,
 \eeq
where $C_L=12/5$ is the Casimir for the link field and $b_B=-5$.
At the scale $m_{v}$,
 the following sfermion masses
are generated by integrating out the link field and the heavy gaugino
\cite{DeSimone:2008gm}:
\begin{equation}
 m_{\tilde{f}}^{2}  = \sum_k  \frac{\alpha_k}{4\pi} C_{\tilde{f},k}
\left(
\frac{2 \alpha_k (\alpha_B-3\alpha_k)}{\alpha_B^2} M_{\tilde{g},B}^2 +
\frac{\alpha_k}{\alpha_B - \alpha_k} m_{v_k}^2
\log \left( 1 + \frac{2 m_L^2}{m_{v_k}^2} \right)
\right)\, . \label{3lferm}
\end{equation}
The $(H_u,H_d)$ soft masses at the scale $m_{v}$ are:
\beq
 m_{H_{u,d}}^{2} =  m_{\tilde{l}}^{2}
+ \lambda_{t,b}^2\frac{\alpha_3}{4\pi}
 \left(\frac{2\alpha_3^2}{\alpha_B^2} M_{\tilde{g},B}^2
 -\frac{\alpha_3}{\alpha_B-\alpha_3}
 \frac{m_{v_3}^2}{2} \log \left( 1+\frac{2 m_L^2}{m_{v_3}^2} \right)
  \right) \, , \label{fo1}
\eeq
where $m_{\tilde{l}}$ is the soft mass of the slepton doublet and
$\lambda_{t,b}$ are the Yukawa couplings of the top and bottom,
respectively.
The MSSM gaugino masses are determined from  $M_{\tilde{g}, B}$:
\begin{equation}
M_{\tilde{g}_k}=  \frac{\alpha_k}{\alpha_B} M_{\tilde{g}, B} \, . \label{fo2}
\end{equation}
Furthermore, rather small trilinear terms are generated
\cite{DeSimone:2008gm}
at the scale $m_v$ by integrating out the heavy gaugino:
\beq
a_t=\frac{\lambda_t}{4\pi}
\left(\frac{16}{3} \alpha_3^2 + 3 \alpha_2^2
+ \frac{13}{15} \alpha_1^2\right)
\frac{M_{\tilde{g},B}}{\alpha_B} \, , \label{fo3}
\eeq
\beq
a_b= \frac{\lambda_b}{4\pi} \left(
\frac{16}{3} \alpha_3^2 + 3 \alpha_2^2
+ \frac{7}{15} \alpha_1^2\right)
\frac{M_{\tilde{g},B}}{\alpha_B} \, ,  \qquad
a_\tau= \frac{\lambda_\tau}{4\pi}
\left(3 \alpha_2^2 + \frac{9}{5} \alpha_1^2\right)
\frac{M_{\tilde{g},B}}{\alpha_B} \, , \nonumber
\eeq
where $\lambda_\tau$ is the Yukawa coupling for the $\tau$.
Below $m_{v}$ the RG equations are solved using SOFTSUSY
\cite{Allanach:2001kg}.
Examples of spectra are shown in  tables \ref{tab3} and
\ref{tab4}.
It turns out that these spectra are rather similar to the ones
considered in the previous section for $m_v \approx \Omega$.

\begin{table}[!htp]
\centering{}\begin{tabular}{|c|c|c|}
\hline
$\Omega$ & $ 1.55\times 10^6 $ & $7.8\times 10^5$  \tabularnewline
$x,y,z$ & 0.8,0.1,1 & 0.98,0.1,1  \tabularnewline
\hline
\hline
$M_{\tilde{g}_B},m_L$ & 3743,46823 & 5429, 27934 \tabularnewline
$M_{\tilde{g}_1},M_{\tilde{g}_2},M_{\tilde{g}_3}$ & 156,274,599 & 224,397,890 \tabularnewline
$m_{\tilde{Q}},m_{\tilde{u}},m_{\tilde{d}}$ & 2634,2518,2507 & 1572,1506,1500 \tabularnewline
$m_{\tilde{L}},m_{\tilde{e}}$ & 843,410 & 491,236 \tabularnewline
$m_{H_u}^2,m_{H_d}^2$ & 736$^{2}$, 835$^{2}$ & 432$^{2}$, 486$^{2}$ \tabularnewline
$\mu,B\mu$ & 956,1271$^{2}$ & 591,828$^{2}$ \tabularnewline
\hline
\hline
$m_{\tilde{g}}$ & 860 & 1161 \tabularnewline
$m{}_{\tilde{\chi}_{0}}$ & 139,279,950,953 & 200,385,587,603 \tabularnewline
$m{}_{\tilde{\chi}_{\pm}}$ & 279,954 & 385,603 \tabularnewline
\hline
$m_{\tilde{u}_{L}},m_{\tilde{d}_{L}}$ & 2674,2675 &  1677,1679 \tabularnewline
$m_{\tilde{u}_{R}},m_{\tilde{d}_{R}}$ & 2554,2544 & 1612,1607 \tabularnewline
$m_{\tilde{t}_{1}},m_{\tilde{t}_{2}}$ & 2356,2575 & 1500,1626 \tabularnewline
$m_{\tilde{b}_{1}},m_{\tilde{b}_{2}}$ & 2523,2573 & 1593,1621 \tabularnewline
\hline
$m_{\tilde{e}_{R}},m_{\tilde{e}_{L}},m_{\tilde{\nu}_{e}}$ & 419,845,841 & 248,504,498 \tabularnewline
$m_{\tilde{\tau}_{1}},m_{\tilde{\tau}_{2}},m_{\tilde{\nu}_{\tau}}$ & 407,843,838 & 238,505,496 \tabularnewline
\hline
$m_{h_{0}}$ & 119 & 116 \tabularnewline
$m_{H_{0}},m_{A_{0}},m_{H_{\pm}}$ & 1203,1203,1206 & 730,730,734 \tabularnewline
\hline
\end{tabular}\caption{Examples of weak scale spectra in the case of
  $p=1$ (two messengers), with $\mu>0$, $\tan \beta = 20$ and
  $\alpha_B^{-1}=4$ at the scale $\Omega$, in the small $m_v$ regime ($m_v\ll \Omega$).
The soft masses $(M_{\tilde{g},B},m_L)$ and
 $(M_{\tilde{g}_k},m_{\tilde{Q},\tilde{u},\tilde{d},\tilde{L},\tilde{e}})$
are given at the scale $m_v=y \Omega$ and $(\mu, B \mu)$ are evaluated
at $m_v$ as well.
 }
\label{tab3}
\end{table}

\begin{table}[!htp]
\centering{}\begin{tabular}{|c|c|c|}
\hline
$\Omega$ & $ 4.1\times 10^5 $ & $ 3.15 \times 10^5 $  \tabularnewline
$x,y,z$ &0.8,0.1,1 & 0.98,0.1,1  \tabularnewline
\hline
\hline
$M_{\tilde{g}_B},m_L$ & 5941,30672 & 13156, 29348 \tabularnewline
$M_{\tilde{g}_1},M_{\tilde{g}_2},M_{\tilde{g}_3}$ & 242,433,994 & 532,950,2169 \tabularnewline
$m_{\tilde{Q}},m_{\tilde{u}},m_{\tilde{d}}$ & 1579,1516,1510 & 1437,1376,1370 \tabularnewline
$m_{\tilde{L}},m_{\tilde{e}}$ & 480,229 & 451,218 \tabularnewline
$m_{H_u}^2,m_{H_d}^2$ & 408$^{2}$, 474$^{2}$ & 408$^{2}$, 448$^{2}$ \tabularnewline
$\mu,B\mu$ & 548,803$^{2}$ & 495,872$^{2}$ \tabularnewline
\hline
\hline
$m_{\tilde{g}}$ & 1254 & 2407 \tabularnewline
$m{}_{\tilde{\chi}_{0}}$ & 219,414,547,572 & 456,497,521,948 \tabularnewline
$m{}_{\tilde{\chi}_{\pm}}$ & 414,571 & 489,948 \tabularnewline
\hline
$m_{\tilde{u}_{L}},m_{\tilde{d}_{L}}$ & 1690,1692 &  1838,1840 \tabularnewline
$m_{\tilde{u}_{R}},m_{\tilde{d}_{R}}$ & 1628,1623 & 1778,1774 \tabularnewline
$m_{\tilde{t}_{1}},m_{\tilde{t}_{2}}$ & 1528,1645 & 1688,1801 \tabularnewline
$m_{\tilde{b}_{1}},m_{\tilde{b}_{2}}$ & 1611,1640 & 1763,1793 \tabularnewline
\hline
$m_{\tilde{e}_{R}},m_{\tilde{e}_{L}},m_{\tilde{\nu}_{e}}$ & 241,495,488 & 245,513,506 \tabularnewline
$m_{\tilde{\tau}_{1}},m_{\tilde{\tau}_{2}},m_{\tilde{\nu}_{\tau}}$ & 232,495,487 & 237,513,505\tabularnewline
\hline
$m_{h_{0}}$ & 116 & 116 \tabularnewline
$m_{H_{0}},m_{A_{0}},m_{H_{\pm}}$ & 694,694,699 & 668,668,673 \tabularnewline
\hline
\end{tabular}\caption{Examples of weak scale spectra in the case of
  $p=6$ couples of messengers,
with $\mu>0$, $\tan \beta = 20$ and $\alpha_B^{-1}=4$ at the scale $\Omega$, in the small
$m_v$ regime ($m_v\ll \Omega$). }
\label{tab4}
\end{table}

It is interesting to compare the magnitude of the two-loop sfermion
soft masses in eq.~(\ref{masisi}) to the three-loop estimation given
by eq.~(\ref{3lferm}).
In the comparison, we just keep the dominant contribution from
eq.~(\ref{3lferm}), arising from the log term which we expand as follows:
\beq
r = \frac{(m_{\tilde{f}}^{2})_{\rm 3-loops}}
{(m_{\tilde{f}}^{2})_{\rm 2-loops}}
\simeq
\frac{6}{5\pi} \frac{\alpha_B^2}{\alpha_B - \alpha_k}
\frac{S(x,\infty,z)}{S(x,y,z)} \ .
\label{estima}
\eeq
This estimate is only valid for $m_{v_k} \gtrsim m_L$.
{}For $\alpha_B^{-1}=4$, one finds
$r\approx 0.16/S(x,y,z)$, showing that the three-loop
contributions are rather important, viz.~the naive loop factor
$\alpha/(4 \pi) \approx 0.01$ gets boosted by a factor of $16$.
The intuitive explanation for this is that the gauge couplings
$g_{A_k},g_B$ are in general stronger than the MSSM gauge coupling
$g_k$.

Throughout the paper we have considered the
$\mathcal{G}_B=SU(5)$ invariant case,
to preserve the unification already present in the MSSM.
However, one might pose the question if
unification is really obtainable in this kind of models,
as we are forced to work with low-scale mediation.
The one-loop beta-function for the gauge group $\mathcal{G}_B$ is IR-free,
conformal and asymptotically free for $p\geq6$, $p=5$ and $p\leq 4$,
respectively. A straightforward analysis shows the following.
For $p\leq 5$, perturbative unification can be obtained
if we give appropriate masses -- of the order of the GUT scale --
to parts of the inert mesons in the embedding theory.
The magnetic dual theory, however, is
not IR-free for $p=5$ which is why we have considered $p=6$ in the
numerical examples. For $p=6$, unification is rather marginal.
If we take equal copies of the messengers, all at the same
scale, we cannot obtain unification due to the coupling $g_B$ running
into a Landau pole before the GUT scale. In order to ameliorate this
problem, we can keep two messengers as in the numerical examples above
and split the remaining five couples by changing the VEV of $\langle
Z\rangle \sim \diag(1,z',z',z',z',z')$, with $z'\gg 1$, giving split
eigenvalues for the rest of the messengers. This results in a small
interval of asymptotically free running and then all the messengers
kick in.
This way we obtain unification of all the couplings of
$\mathcal{G}_A$ and avoid the Landau pole problem for $g_B$,
though we get a somewhat strongly coupled unification,
i.e.~$\alpha_{\rm GUT}$ of order $1$.

\newpage
\subsubsection{$\mathcal{G}_B= U(1) \times SU(2) \times SU(3)$} \label{sec123}

In this section we discuss the case in which the gauge group
$\mathcal{G}_B$
is $ U(1) \times SU(2) \times SU(3)$, instead of $SU(5)$,
and each of the factors has a {\it different} gauge coupling $g_{B_k}$.
The charges of the link fields are:
\begin{center}
\begin{tabular}{c|c@{}c@{}c@{}c@{}c|c@{}c@{}c@{}c@{}c}
 & $ U(1)_A$ & $\times$ & $SU(2)_A$ & $\times$ &  $SU(3)_A $ &
$U(1)_B$ & $\times$ & $SU(2)_B$ & $\times$ & $SU(3)_B$  \\
\hline
$L_2$ & $-\frac{3}{2\sqrt{15}}$ && ${\bf 2}$ && ${\bf 1}$ &
$\frac{3}{2\sqrt{15}}$ && ${\bf 2}$ && ${\bf 1}$ \\
$\tilde{L}_2$ & $\frac{3}{2\sqrt{15}}$ && ${\bf 2}$ && ${\bf 1}$ &
$-\frac{3}{2\sqrt{15}}$ && ${\bf 2}$ && ${\bf 1}$ \\
$L_3$ & $\frac{1}{\sqrt{15}}$ && ${\bf 1}$ && ${\bf 3}$ &
$-\frac{1}{\sqrt{15}}$ && ${\bf 1}$ && $\bar{\bf 3}$ \\
$\tilde{L}_3$ & $-\frac{1}{\sqrt{15}}$ && ${\bf 1}$ && $\bar{\bf 3}$ &
$\frac{1}{\sqrt{15}}$ && ${\bf 1}$ && ${\bf 3}$
\end{tabular}
\end{center}
At the scale $\Omega$ the sfermion masses are taken to be zero,
while  the gaugino of the gauge group $\mathcal{G}_{B_k}$
gets a soft mass $M_{\tilde{g}, B_k}$, which can be computed from the
expression (\ref{gauma}, \ref{maga}), with $\alpha_k$ replaced
 by $\alpha_{B_k}$.
The link fields $(L_2,L_3)$ obtain the following soft masses from
gauge mediation:
\begin{align}
m_{L_2}^2 &= 2 \, p
\left(\frac{f}{\Omega}\right)^2 S(x,\infty,z)
\left(  \frac{3}{4}  \left(\frac{\alpha_{B_2}}{4\pi}\right)^2 +
  \frac{3}{20}   \left(  \frac{\alpha_{B_1}}{4\pi}\right)^2 \right) \,
  , \\
m_{L_3}^2 &= 2 \, p
\left(\frac{f}{\Omega}\right)^2 S(x,\infty,z)
\left(  \frac{4}{3} \left(\frac{\alpha_{B_3}}{4\pi}\right)^2  +
\frac{1}{15}  \left(\frac{\alpha_{B_1}}{4\pi}\right)^2  \right) \,
. \nonumber
\end{align}
{}From the scale $\Omega$ to $m_{v}$, the renormalization group
equations are \cite{martin-review}:
\begin{align}
\frac{d \, M_{\tilde{g}, B_k} }{d (\log \mu)} &=
\frac{b_{B_k}}{8 \pi^2} g^2_{B_k}  M_{\tilde{g},B_k } \, , \qquad
b_{B_k} = \left( \frac{12}{5} , -2, -3 \right) \, , \non
\frac{d \, m^2_{L,2} }{d (\log \mu)} &=
- \frac{3}{4} \frac{1}{2\pi^2} g_{B_2}^2 M_{\tilde{g},B_2}^2
-\frac{3}{20} \frac{1}{2\pi^2} g_{B_1}^2 M_{\tilde{g},B_1}^2
 \, , \\
\frac{d \, m^2_{L,3} }{d (\log \mu)} &=
- \frac{4}{3} \frac{1}{2\pi^2} g_{B_3}^2 M_{\tilde{g},B_3}^2
-\frac{1}{15} \frac{1}{2\pi^2} g_{B_1}^2 M_{\tilde{g},B_1}^2
 \, . \nonumber
\end{align}
At the scale $m_{v}$,
 the following sfermion masses
are generated by integrating out the link field and the heavy gaugino:
\begin{equation}
 m_{\tilde{f}}^{2}  = \sum_k  \frac{\alpha_k}{4\pi} C_{\tilde{f},k}
\left(
\frac{2 \alpha_k (\alpha_{B_k}-3\alpha_k)}{\alpha_{B,k}^2} M_{\tilde{g},{B_k}}^2 +
\frac{\alpha_k}{\alpha_{B_k} - \alpha_k} m_{v_k}^2
\log \left( 1 + \frac{2 m_{L_k}^2}{m_{v_k}^2} \right)
\right)\, , \nonumber
\end{equation}
where
\beq
m_{L_1}^2=\frac{2 m_{L_3}^2
+ 3 m_{L_2}^2  }{5} \, .
\eeq
The Higgs soft masses are given by eq.~(\ref{fo1}) with $\alpha_B$
replaced by $\alpha_{B,3}$.
The MSSM gaugino masses are determined from  $M_{\tilde{g},B_k}$:
\begin{equation}
M_{\tilde{g}_k}=  \frac{\alpha_k}{\alpha_{B_k}} M_{\tilde{g}, B_k} \, .
\end{equation}
The trilinear scalar couplings at the scale $m_v$ are now:
\begin{align}
a_t &= \frac{\lambda_t}{4\pi}
\left(\frac{16}{3} \frac{\alpha_3^2}{\alpha_{B_3}}  M_{\tilde{g},B_3}
 + 3 \frac{\alpha_2^2}{\alpha_{B_2}}  M_{\tilde{g},B_2}
+ \frac{13}{15} \frac{ \alpha_1^2}{\alpha_{B_1}}   M_{\tilde{g},B_1}  \right)
\, , \non
a_b &= \frac{\lambda_b}{4\pi}
\left(\frac{16}{3} \frac{\alpha_3^2}{\alpha_{B_3}}  M_{\tilde{g},B_3}
 + 3 \frac{\alpha_2^2}{\alpha_{B_2}}  M_{\tilde{g},B_2}
+ \frac{7}{15} \frac{ \alpha_1^2}{\alpha_{B_1}}   M_{\tilde{g},B_1}  \right)
\, , \\
a_\tau &= \frac{\lambda_\tau}{4\pi}
\left( 3 \frac{\alpha_2^2}{\alpha_{B_2}}  M_{\tilde{g},B_2}
+ \frac{9}{5} \frac{ \alpha_1^2}{\alpha_{B_1}}   M_{\tilde{g},B_1}  \right)
\, . \nonumber
\end{align}

The Higgs mass gets a contribution
which is absent in MSSM, due to the fact that the D-term of the heavy
gauge boson does not decouple completely if there is supersymmetry
breaking.
The usual MSSM D-terms are modified to \cite{DeSimone:2008gm}:
\begin{align}
V_{D}= \frac{g_2^2 (1+\Delta_2)}{8} \left|H_u^\dagger \sigma^a H_u + H_d^\dagger \sigma^a H_d\right|^2
+\frac{3}{5} \frac{g_1^2 (1+\Delta_1)}{8} \left|H_u^\dagger H_u - H_d^\dagger H_d\right|^2 \, ,
\end{align}
where $\sigma^a$ are the Pauli matrices and $\Delta_k$ are given by:
\beq
\Delta_k = \frac{\alpha_k}{\alpha_{B_k} - \alpha_k} \frac{2 m_{L_k}^2}{m_{v_k}^2+ 2 m_{L_k}^2} \, .
\label{dkdk}
\eeq
In the presence of $\Delta_{1,2}$, the usual bound $m_{h_0} < m_Z$ at tree level
(which is saturated at large $\tan \beta$) is replaced by
 \cite{Batra:2003nj,Maloney:2004rc,Craig:2011ev,Craig:2011yk}:
\beq
m_{h_0}^2 < \tilde{m}^2 \, , \qquad
\tilde{m}^2= \frac{ \frac{3}{5} g_1^2(1+ \Delta_1) +g_2^2 (1+\Delta_2)}{2} \, v_h^2 \, ,
\qquad v_h = 174 \, {\rm GeV}\, .
\label{mmvvhh}
\eeq
If $\Delta_{1,2}$ is of order one, this contribution is quite useful
to ameliorate the little hierarchy problem.
{}For the concrete case considered in tables \ref{tab1}--\ref{tab4}, this
is negligible because we considered the $SU(5)_B$ invariant
coupling $\alpha_B^{-1}=4$ at the messenger scale, which is large
compared to $\alpha_{1,2}$. Choosing for example $\alpha_B^{-1}=7$
would give an extra contribution to $m_{h_0}$ of up to about $10 \,
{\rm GeV}$. However, for $\alpha_B^{-1} \gtrsim 4$ we will have to give
up unification anyway (i.e.~$\alpha_{\rm GUT}$ will become strongly
coupled) and hence there is no motivation for keeping the
$SU(5)$ invariant couplings.
So if instead we consider the $U(1)\times SU(2) \times SU(3)$ case
and choose e.g.~$\alpha_{B_{i}} \approx \alpha_{A_i}$,
the extra contribution to $m_{h_0}$ can be up to about
$30 \, {\rm GeV}$.

We can use SOFTSUSY to compute the mass of all
the particles in the spectrum with the exception of the Higgses,
which we compute at tree level from the soft masses at the weak scale.
The tree-level masses of the Higgses are:
\begin{align}
m^2_{A_0}&=\frac{m^2_{H_u} -m^2_{H_d}}{\cos 2 \beta} - \tilde{m}^2 \, , \qquad
m^2_{H_\pm}=\frac{m^2_{H_u} -m^2_{H_d}}{\cos 2 \beta} -
 \frac{ \frac{3}{5} g_1^2(1+ \Delta_1) }{2} \, v_h^2 \, , \\
m^2_{h_0,H_0}&=\frac{1}{2} \left(m^2_{A_0} + \tilde{m}^2 \mp
\sqrt{(m^2_{A_0}-\tilde{m}^2)^2+4 \tilde{m}^2 m^2_{A_0} \sin^2 2
  \beta}  \right) \, , \nonumber
\end{align}
where $\tilde{m}$ and $v_h$ are given in eq. (\ref{mmvvhh}).
{}For $h_0$ we add the one-loop correction \cite{martin-review}:
\beq
\Delta_{m^2_{h_0}} = \frac{3}{4 \pi^2} (\cos^2 \alpha) \, \lambda_t^2 m_t^2
 \log \frac{m_{\tilde{t}_1}  m_{\tilde{t}_2}}{m_t^2} \,,
\eeq
where $\alpha$ is the Higgs mixing angle, which can be obtained from the tree-level masses as follows:
\beq
\sin 2 \alpha = -\frac{m^2_{H_0}+m^2_{h_0}}{m^2_{H_0}-m^2_{h_0}} \sin 2 \beta \, .
\eeq
Examples of spectra in the small $m_v$ regime
(to allow large $\Delta_{1,2}$)
are shown in table \ref{tab5}.
The Higgs $h_0$ can have a mass near $140 \, {\rm GeV}$,
with a stop mass near $900 \, {\rm GeV}$.

 \begin{table}[!htp]
\centering{}\begin{tabular}{|c|c|c|}
\hline
$\Omega$ & $ 3.8 \times 10^5 $ & $ 2.5\times 10^5$  \tabularnewline
$x,z$ &0.8,1 & 0.98,1  \tabularnewline
$y_1,y_2,y_3$ &0.010,0.014,0.022 & 0.010,0.014,0.022  \tabularnewline
\hline
\hline
$m_{L_2},m_{L_3}$ & 4007, 10867 & 3111,8639 \tabularnewline
$M_{\tilde{g}_1},M_{\tilde{g}_2},M_{\tilde{g}_3}$ & 216,400,1004 & 406,753,1897 \tabularnewline
$m_{\tilde{Q}},m_{\tilde{u}},m_{\tilde{d}}$ & 979,964,959 & 663,651,648\tabularnewline
$m_{\tilde{L}},m_{\tilde{e}}$ & 222,168 & 155,117 \tabularnewline
 $m_{H_u}^2,m_{H_d}^2$ & 162$^{2}$, 218$^{2}$ & 124$^{2}$, 153$^{2}$ \tabularnewline
$\mu,B\mu$ & 292,442$^{2}$ & 181,362$^{2}$ \tabularnewline
\hline
\hline
$m_{\tilde{g}}$ & 1132 & 1932 \tabularnewline
$m{}_{\tilde{\chi}_{0}}$ & 193,279,298,430 & 172,187,393,754\tabularnewline
$m{}_{\tilde{\chi}_{\pm}}$ & 272,429 & 180,754\tabularnewline
\hline
$m_{\tilde{u}_{L}},m_{\tilde{d}_{L}}$ & 1086,1089 &  908,911\tabularnewline
$m_{\tilde{u}_{R}},m_{\tilde{d}_{R}}$ & 1070,1066 & 891,889 \tabularnewline
$m_{\tilde{t}_{1}},m_{\tilde{t}_{2}}$ & 1035,1080 & 867,909 \tabularnewline
$m_{\tilde{b}_{1}},m_{\tilde{b}_{2}}$ & 1057,1071 & 883,897 \tabularnewline
\hline
$m_{\tilde{e}_{R}},m_{\tilde{e}_{L}},m_{\tilde{\nu}_{e}}$ & 178,244,230 & 137,213,197 \tabularnewline
$m_{\tilde{\tau}_{1}},m_{\tilde{\tau}_{2}},m_{\tilde{\nu}_{\tau}}$ & 167,250,229 & 131,216,197 \tabularnewline
\hline
$m_{h_{0}}$ & 139 & 139 \tabularnewline
$m_{H_{0}},m_{A_{0}},m_{H_{\pm}}$ & 488, 487, 497 & 460, 460, 470\tabularnewline
$\Delta_1,\Delta_2$ & 0.88,0.53 & 0.95, 0.67 \tabularnewline
\hline
\end{tabular}\caption{Examples of weak scale spectra in the case of
  $p=6$ couples of messengers,
with $\mu>0$, $\tan \beta = 20$ and $\alpha_{B_k}^{-1}=\alpha_{A_k}^{-1}$ at the scale $m_v$,
in the small $m_v$ regime ($m_v\ll \Omega$).}
\label{tab5}
\end{table}

\section{Discussion} \label{sec:conclusions}

In this work, we inspected the sparticle spectrum
in direct gaugino mediation \cite{Green:2010ww}.
The main result is the following.
We found a relatively reasonable spectrum
already in the hybrid case,
when the various scales in the problem --
the messenger scale $\Omega$,
the effective SUSY-breaking scale $f/\Omega$,
the R-symmetry breaking scale $\omega$
and the Higgsing scale $m_v$ --
are comparable.
Concretely, from table \ref{tab2},
we see that the gluino mass is near the TeV --
above the current LHC limits,
while the stop mass is in the 1--2 TeV range.
Intriguingly, when the SUSY-breaking scale
is sufficiently close to the messenger scale,
the NLSP might be a stau,
followed by a right-handed selectron
with a comparable mass.

{}From table \ref{tab4}, we see that
the superpartner spectrum in the gaugino mediation regime --
when the Higgsing scale $m_v$ is much smaller
than the messenger scale $\Omega$ --
is rather similar to the one in the hybrid case.
This is due to the large three-loop
contributions to the soft scalar masses in this case.
It would be interesting to investigate also the intermediate regime,
where the two and three-loop contributions to the soft sfermion
masses are comparable.

Finally, if we do not require unification --
e.g. when both gauge groups in figure \ref{qui} are $SU(3)\times SU(2)\times U(1)$
with comparable couplings -- and deep in the gaugino mediation regime,
$m_v\ll\Omega$, we find significant contribution to the Higgs potential,
which is useful to ameliorate the little hierarchy problem.
Concretely, in the second column of table \ref{tab5},
the stop mass is near 900 GeV and
the SM Higgs mass is about 140 GeV.

\subsection*{Acknowledgments}

We are grateful to Shmuel Elitzur, Andrey Katz, Boaz Keren-Zur and Zohar Komargodski
for valuable discussions.
This work was supported in part
by the BSF -- American-Israel Bi-National Science Foundation,
and by a center of excellence supported by the Israel Science Foundation
(grant number 1665/10).
SBG gratefully acknowledges a Golda Meir post-doctoral fellowship.

\end{document}